\documentclass[twocolumn,preprintnumbers,amsmath,amssymb]{revtex4}
\def\ket{\rangle}
\def\bra{\langle}

\usepackage{graphicx}

\usepackage{amsmath}
\begin{document}
\preprint{GLL/07-24}
\title{Duality  and Recycling  Computing in Quantum Computers}
\author{Gui Lu Long and  Yang Liu}

%
\affiliation{$^1$ Key Laboratory of Atomic and Molecular NanoSciencs and Department of
Physics, Tsinghua University, Beijing 100084, P. R. China\\
$^2$ Tsinghua National Laboratory for Information Science and
Technology, Beijing 100084, P. R. China}
\date{\today}



\maketitle

Quantum computer possesses quantum parallelism and offers great
computing power over classical computer \cite{er1,er2}. As is
well-know, a moving quantum object passing through a double-slit
exhibits particle wave duality. A quantum computer is static and
lacks this duality property. The recently proposed duality computer
has exploited this particle wave duality property, and it may offer
additional computing power \cite{r1}. Simply put it, a duality
computer is a moving quantum computer passing through a double-slit.
A duality computer  offers the capability to perform separate
operations on the sub-waves
 coming out of the different slits, in the so-called duality parallelism. Here we show that
 an $n$-dubit duality computer can be modeled by
an $(n+1)$-qubit quantum computer. In a duality mode, computing
operations are not necessarily unitary. A $n$-qubit quantum computer
can be used as an $n$-bit reversible classical computer and is
energy efficient. Our result further enables a $(n+1)$-qubit quantum
computer to run classical algorithms in a $O(2^n)$-bit classical
computer. The duality mode provides a natural link between classical
computing and quantum computing. Here we also propose a recycling
computing mode in which a quantum computer will continue to compute
until the result is obtained. These two modes provide new tool for
algorithm design. A search algorithm for the unsorted database
search problem is designed.

 In a duality computer, there exist two new
computing gates in addition to the usual universal gates for quantum
computers \cite{r1}. One such  gate is  the quantum wave divider
(QWD), and a double-slit is an example of such gate. Suppose there
is a complex Hilbert space ${H}$, a QWD reproduces copies of the
wave function with attenuated coefficient in $m$ direct summed
Hilbert spaces ${H}^{\oplus^m}=\sum_{i=1}^m \oplus H_i$, namely
\begin{eqnarray} D_p (|\psi\ket)=\sum_{i=1,m}\oplus p_i|\psi\ket_i,
\end{eqnarray}
where $\sum_ip_i=1$ and $p_i$ is the strength of the sub-wave in the
$i$-th slit, $m$ is the number of slits. Another duality gate is the
reverse of QWD, the quantum wave combiner (QWC), and it combines the
sub-waves in $H^{\oplus^m}$ into a single Hilbert space $H$,
\begin{eqnarray}
C(p_1|\psi_1\ket\oplus\cdots\oplus
p_m|\psi_m\ket)=\sum_i^mp_i|\psi_i\ket.
\end{eqnarray}
A duality computing process is described in the following
\begin{eqnarray}
|\psi\ket\rightarrow\sum_i^m \oplus p_i|\psi\ket_i\rightarrow
\sum_i^m \oplus p_iU_i|\psi\ket_i\rightarrow\sum_i^m p_i
U_i|\psi\ket.\label{dc}
\end{eqnarray}
The general quantum gate $\sum_i p_i U_i$, or duality gate, is no
longer unitary. As a result, the final wave function in Eq.
(\ref{dc}) is only part of a wave function.
 The power of duality computer depends sensitively on the result of
 measurement of part of a wave function.
   Three possibilities have been
 suggested \cite{r1}:
 1) one will get a result immediately as if  a whole wave
 function were measured; 2) one will get a result but with a longer time;
 3) one sometimes gets a
 result  and sometimes one does not. In the first two
 scenarios, a duality computer could solve NP-complete problems
 in polynomial time
 \cite{r1,r2}.
  The mathematical description of duality computer with the first two scenarios
 has been given recently
\cite{r3,r4}.  In this work, we assume  case 3, which comes out from
the measurement postulate of quantum mechanics naturally.

A symmetric  2-routes duality computer has $p_1=p_2=1/2$.  The
complete wave function of a duality computer is
$|\psi\rangle=|\varphi\rangle|\kappa\rangle$ where $|\varphi\rangle$
is the internal state and $|\kappa\rangle$ is the center of mass
translational motion wave function. When a QWD operation is
performed, it changes the state to
\begin{equation}
|\psi\rangle^\prime=\sum^{2}_{i=1}\oplus
\frac{1}{2}|\varphi\rangle|\kappa_i\rangle.\label{e1}
\end{equation}
One then performs different gate operations on the sub-waves,
\begin{equation}
|\psi\rangle^{\prime\prime}=\frac{1}{2}\sum^{2}_{i=1}\oplus
U_i|\varphi\rangle|\kappa_i\rangle.\label{e2}
\end{equation}
Then a QWC operation is performed and changes the state to
\begin{equation}
|\psi\rangle_f=(p_1U_1+p_2U_2)|\varphi\rangle|\kappa\rangle.\label{e3}
\end{equation}
A final measurement is performed on $|\psi\rangle_f$ to read-out the
outcome. Under the 3rd assumption about partial measurement, the
wave function should not renormalized, and a result to a conditional
measurement is obtained only with some probability. Whereas in the
other two scenarios, a result to a partial measurement is always
obtained and the wave function should be renormalized.

The fundamental difference between duality computer and quantum
computer is that duality gates need not be unitary. A quantum
computer can not perform $U_1+U_2$, it can only perform $U_1U_2$
operation.

\begin{figure}[here]
\begin{center}
\includegraphics[width=8cm]{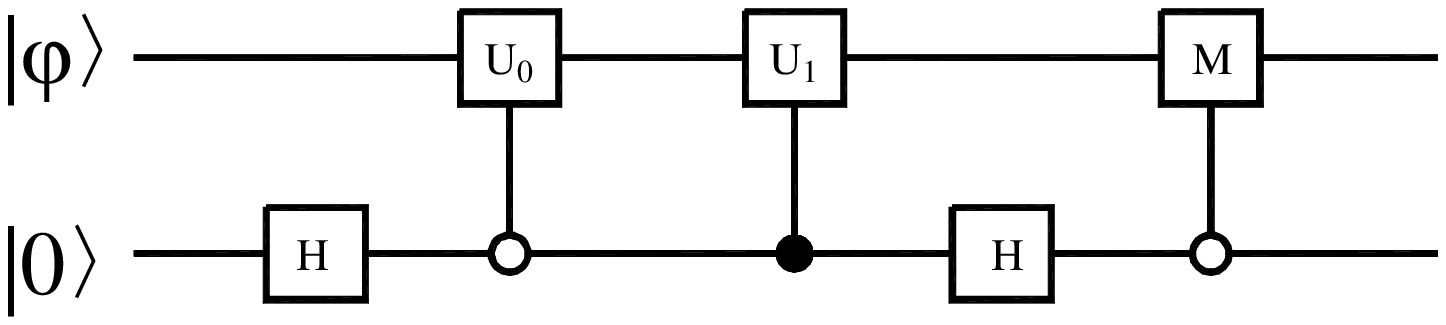}
\caption{An $n$-dubit duality computer can be simulated by an
$(n+1)$ qubit quantum computer. The Walsch-Hadamard gates can be
replaced by other unitary gates to simulate more complicated slits.
One can also use more qubits as auxiliary to simulate more slits.
}\label{f1}
\end{center}
\end{figure}

{\bf Duality  Mode in a Quantum Computer}---Now we give a quantum
computer simulation of the duality computer. To simulate an
$n$-dubit duality computer, we need an $(n+1)$-qubit quantum
computer, and one  qubit is used as auxiliary qubit and $n$ qubits
are used as work qubits. Let's make the following correspondence
\begin{eqnarray}
&&|\varphi\rangle|\kappa_u\rangle\leftrightarrow |\varphi\rangle|0\rangle,\nonumber\\
&&|\varphi\rangle|\kappa_d\rangle\leftrightarrow |\varphi\rangle|1\rangle, \label{e5}
\end{eqnarray}
namely, when the auxiliary qubit is in $|0\rangle(|1\rangle)$, it
resembles a duality computer sub-wave from the upper (lower) slit.
We ascribe the initial and final wave function of the duality
computer to the wave function when the auxiliary is in state
$|0\rangle$. Thus, before the QWD, the state of duality computer is
$|\varphi\rangle|0\rangle$, after the QWD, the wave-function becomes
\begin{equation}
|\varphi\rangle\frac{|0\rangle+|1\rangle}{\sqrt{2}},\label{e6}
\end{equation}
namely, the QWD is simulated by a Walsch-Hadamard transformation in
the auxiliary qubit. Thus the wave-function describes the $n$-bit
quantum computer simultaneously in two slits, $|0\rangle$ and
$|1\rangle$. Different gate operations  on different routes can be
simulated using conditional gates, as shown in Fig. \ref{f1}.

Then we have
\begin{equation}
\frac{U_0|\varphi\rangle|0\rangle+U_1|\varphi\rangle|1\rangle}{\sqrt{2}}.\label{e7}
\end{equation}
The quantum combiner operation is again a Walsch-Hadamard
transformation, after which the wave-function becomes
\begin{equation}
\frac{U_0+U_1}{2}|\varphi\rangle|0\rangle+\frac{U_0-U_1}{2}|\varphi\rangle|1\rangle.\label{e8}
\end{equation}
We make a measurement on condition that the auxiliary qubit is
$|0\rangle$. The probability to obtain a result is the square of the
norm of the wave function in $|0\ket$ state, namely $P_0=\bra
\varphi|(U_0+U_1)^{\dagger}(U_0+U_1)|\varphi\ket/4$. With
probability $1-P_0$, the conditional measurement will not obtain a
result, and if this occurs, the state of the wave function is
collapsed in a normalized state $N(U_0-U_1)|\varphi\ket|1\ket$. In
this simulation, the conditional measurement simulates a measurement
on a part of a wave function. The result is clear from the basic
postulate in quantum mechanics.

The scheme can be generalized by replacing the two Walsch-Hadamard
gates with other unitary gates to simulate more complicated slits,
such as asymmetric slits, slits with different phases and so on. By
adding more auxiliary qubits, one can simulate a duality computer
with more slits.

We now give the mathematical description for such duality mode. Some
of the result in Ref. \cite{r3} can be used directly, though the
results regarding QWD and QWC need  modification. Denoting the set
of duality gates on $H$ as $\mathcal{G}(H)$, there are four theorems
and corollaries regarding duality gates in Ref. \cite{r3} and we
directly copy them here. For proof of these results, see Ref.
\cite{r3}.

{\bf Theorem 1.} The identity $I_H$ is an extreme point of $\mathcal
{G}(H)$.

{\bf Corollary 2.} The extreme points of $\mathcal{G}(H)$ are
precisely the unitary operators in $H$.

The theorem and corollary tell us that the duality gate is unitary
only when $U_0=U_1$, and in particular, when $U_0=U_1=I_H$, the
duality gate is the identity gate.

Let $\mathcal{B}(H)$ be the set of bounded linear operators on $H$
and let $\mathbb{R}^+\mathcal{G}(H)$ be the positive cone generated
by $\mathcal{G}(H)$, that is
\begin{eqnarray}
\mathbb{R}^+\mathcal{G}(H)=\left\{\alpha A: A\in
\mathcal{G}(H),\alpha\geq 0\right\}.
\end{eqnarray}
Then the next theorem tells that any operator on $H$ can be
simulated.

{\bf Theorem 3}. If dim $H<\infty$, then
$\mathcal{B}(H)=\mathbb{R}^+\mathcal{G}(H)$.

The next corollary is about normal operators.

{\bf Corollary 4}. If dim $H<\infty$, then $A\in \mathcal{B}(H)$ is
normal if and only if $A=\alpha\sum_i p_i U_i$ where $\alpha\geq 0$,
$p_i>0$, $\sum_i p_i=1$ and $U_i$ are unitary operators that
mutually commute.

We now expound the significance of duality mode in a quantum
computer. First, an $(n+1)$-qubit quantum computer can simulate an
$n$-dubit duality computer. This allows a quantum computer to
perform any operation in the $n$-qubit Hilbert space. Hence,
classical algorithms can be translated into quantum algorithms in
this duality  mode. This is significant because this is not a direct
use of a quantum computer as a classical computer using a qubit as a
classical bit, and the qubit resource in duality mode is
exponentially small. For instance for an unsorted database search
problem with $N=2^n$ items, a classical computer needs at least $n
2^n=N n$ bits to express the database. However in a quantum computer
running in duality mode, it needs only $(n+1)$ qubits to express and
manipulate the database. This also provides a natural bridge between
classical and quantum computing. It was suggested that future
quantum computer may be used as special-purpose processor to perform
task that needs quantum acceleration, and then the result is
returned to a classical computer for further processing. Using this
duality mode, one may perform both tasks in a quantum computer.

Secondly, the duality mode has provided a new avenue for algorithm
design. Let's study the unsorted database search problem \cite{r5}.
Quantum algorithms find a marked item from an unsorted database with
$O(\sqrt{N})$ steps \cite{er2,r7,r8,r9}.  However these quantum
algorithms are not fixed point search algorithm, the success
probability is a periodic function of the number of searching steps.
A recent fixed point quantum search algorithm uses $O(3^n)$ number
of queries in a quantum computer \cite{r10}. Though the number of
queries is more than $O(2^n)$ in a classical computer, it is still
advantageous because it is an algorithm in a quantum computer using
only $n$-qubit. Here we give a duality mode algorithm that uses only
$O(2^n)$ steps while being still a quantum computer algorithm. This
is a modified duality algorithm in Ref. \cite{r1}.
 One starts from the
evenly distributed state, and switch onto the duality mode to give
\begin{equation}
\frac{1}{\sqrt{2N}}\sum^{N-1}_{i=0}|i\rangle(|0\rangle+|1\rangle).
\label{e9}
\end{equation}
Performing on the upper slit the following gate operation
\begin{equation}
\frac{1}{\sqrt{2N}}(|0\rangle+\ldots+|\tau\rangle+\ldots)\rightarrow
\frac{1}{\sqrt{2N}}(-|0\rangle-\ldots+|\tau\rangle-\ldots),\label{e10}
\end{equation}
and leaves the lower-slit sub-wave unchanged. Then recombine them
through a Hadamard gate, we obtain
\begin{equation}
\frac{1}{\sqrt{N}}|\tau\rangle|0\rangle+\frac{1}{\sqrt{N}}
\sum_{i\neq\tau}|i\rangle|1\rangle.\label{e11}
\end{equation}
Then conditionally measure  on the auxiliary state being in
$|0\rangle$, one obtains $\tau$ with probability $1/N$. Conditional
measurement is simply achieved by measuring the auxiliary qubit. If
it is 0, then measure the working $n$ qubit to read out $\tau$. If
it is 1, repeat the process again.  Repeating this process
$O(N=2^n)$ number of times, one will get the desired result.

\begin{figure}[here]
\begin{center}
\includegraphics[width=8cm]{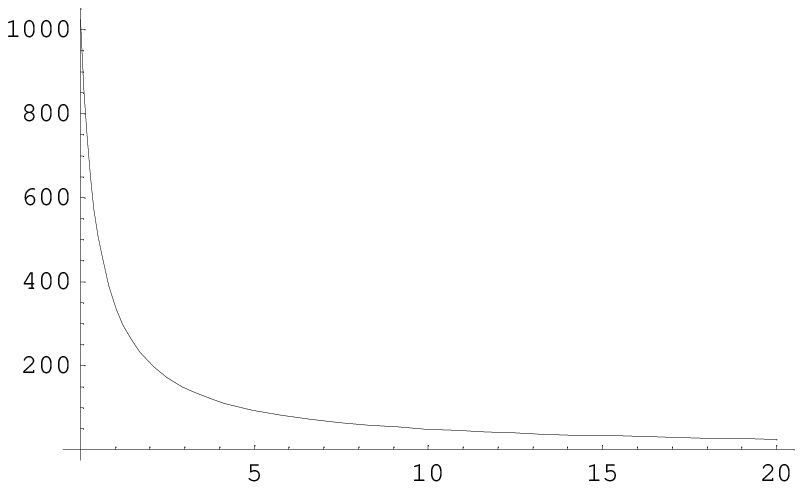}
\caption{The repetition number  versus $j$ for
$N=2^{10}$.}\label{f2}
\end{center}
\end{figure}

The algorithm can be further speeded up by first performing the
quantum amplitude amplification a number of times before switching
to duality mode. After $j$ iteration, the wave function becomes
\begin{eqnarray}
|\psi_j\ket=\sin((2j+1)\beta)|\tau\ket+\cos((2j+1)\beta)|c\ket,
\label{e12}\end{eqnarray} where
\begin{eqnarray}
|c\ket=\sqrt{1\over N-1}\sum_{i\ne \tau}|i\ket.\label{e13} \end{eqnarray} Then
switching to the duality mode, it gives for the wave function in the upper slit
\begin{eqnarray} |\psi_u\ket=\sin((2j+1)\beta)|\tau\ket. \label{e14}\end{eqnarray}
Conditionally measure it one gets $\tau$ with probability
$\sin^2[(2j+1)\beta]$. Repeating this $O(1/\sin((2j+1)\beta)$ times,
the marked state will be found. When $j$ is small, the number of
repetitions is about $N/(2j+1)$. When $j$ approaches
$\pi\sqrt{N}/4$, it finds the marked state with only single query. A
schematic plot for $N=2^{10}$ is given in Fig. \ref{f2}. When one
knows the number of marked state in an unsorted database, one can
optimize the number of repetition. If one does not know this
information, one can simply switch to the duality mode, and repeat
the process until a result is obtained in the conditional
measurement. This algorithm works also for an arbitrary database
where the coefficient of each item is arbitrary. Using the recycling
mode, the repetition process is performed automatically.

Thirdly, this formulation of duality computer has provided a way for
error correction in duality computer. Quantum error correction has
been solved successfully in quantum computer \cite{r11}, this
analogy easily gives the scheme of error correction in a duality
computer. All the good quantum error codes, can be translated into
duality computer.

\begin{figure}[here]
\begin{center}
\includegraphics[angle=0,width=8cm]{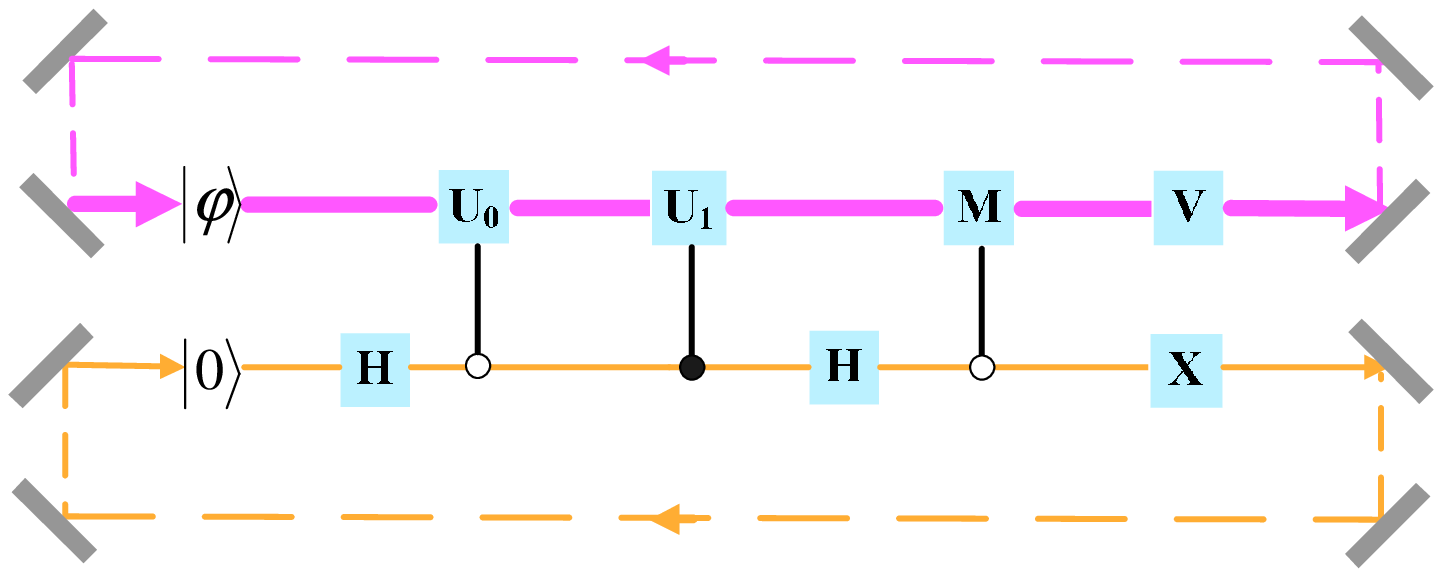}
\caption{Schematic illustration of a recycling duality computation.
After the auxiliary-qubit-conditioned measurement, if a result is
obtained, the calculation is completed and the process is stopped.
If no result is obtained, it destroys the state in $|0\ket$
auxiliary qubit, and leaves a state where the auxiliary is in state
$|1\ket$. The state is restored to the input state by a unitary
operation $V$ and guided to the input end. The calculation process
is repeated again and again until the final result is obtained.
}\label{f3}
\end{center}
\end{figure}

{\bf Recycling quantum computing}--- In Eq.(\ref{e11}), the
probability of obtaining a result in a conditional measurement is
small. However it is different from an evenly distributed state
$\frac{1}{\sqrt{N}}\sum_i|i\rangle$. When one measures the evenly
distributed state, one always obtains a result, say $|x\rangle$,
however the result to be $\tau$ has only a probability of $1/N$,
afterwards the state collapses to the eigenstate of that measured
eigenvalue. However, in Eq. (\ref{e11}), after the measurement, one
has two possibilities : 1) The state collapses into state
$|\tau\ket$ with probability $1/N$; 2) No result is obtained,
however the state has {\bf collapsed out} from (\ref{e11}), and the
state becomes
$$\frac{1}{\sqrt{N-1}}\sum_{i\ne \tau}|i\rangle|1\rangle.$$
This happens with probability $(N-1)/N$. This state can be reused
again as input after  flipping the auxiliary qubit and apply a
recovering unitary operation to the original input state. Then the
calculating process is recycled again and again until a final result
is obtained.  Using this observation, we propose a recycling quantum
computing mode as shown in Fig. \ref{f3}.

Namely, before the conditional measurement, suppose the
wave-function of the duality computer is
\begin{equation}
\frac{U_0+U_1}{2}|\varphi\rangle|0\rangle
+\frac{U_0-U_1}{2}|\varphi\rangle|1\rangle.\nonumber
\end{equation}
After the conditional-measurement, if a result is obtained,
the wave-function is
collapsed, and the $(U_0+U_1)|\varphi\rangle$ result is read out.
 If no-result is
obtained, then the state in $|0\rangle$ collapses out,
and the wave-function becomes
\begin{equation}
|\psi'\ket=N'\frac{U_0-U_1}{2} |\varphi\rangle|1\rangle,\nonumber
\end{equation}
where $N'$ is a renormalization constant.   Then performing a
unitary operation $V$ on the $n$ qubits,  the initial input state is
restored. Meanwhile  flipping $|1\rangle$ to $|0\rangle$ in the
auxiliary qubit. The $(n+1)$ qubits are guided  into the beginning
of the circuit, as input state. The process will continue until a
result is read-out in the conditional measurement device.

This simulation of duality computer also provides a relativity view
regarding quantum computer. In a duality computer, a double-slit is
located statically, and a quantum computer is moving. However in the
duality mode, a quantum computer is static, and the double-slit, the
auxiliary qubit, changes to simulate the motion of the double-slit.
To make a quantum computer moving is very difficult, however it is
much easier to add one additional qubit to an $n$-qubit quantum
computer.

 In summary we have given a quantum
computer realization of the duality computer. This realization
itself serves as a new mode of quantum computing, the duality mode.
This provides a way to run classical algorithm in quantum computers
using much reduced qubit resources. It also provides a method for
error correction in a duality computer. With conditional
measurement, the recycling quantum computing mode is also proposed.
These two modes  provide new ways and flexibility in quantum
algorithm designs.

{\bf Acknowledgement}

Helpful discussions with Prof. Dieter Suter and Dr Qing-Yu Cai are
gratefully acknowledged. This work is supported by the National
Fundamental Research Program Grant No. 2006CB921106, China National
Natural Science Foundation Grant Nos. 10325521, 60433050.



\begin{thebibliography}{99}

\bibitem{er1} Shor, P W. Algorithms for quantum computation:
discrete logarithms and factoring. In: Proceedings of the Symposium
on the Foundations of computer Science. New York: IEEE Computer
Society Press, 1994, 124-134.

\bibitem{er2} Grover, L K, A fast quantum mechanical algorithm for
database search. In: Proceedings of 28th Annual ACM Symposium on
Theory of Computing. New York: ACM,  212-219 (1996).


\bibitem{r1} Long, G L, {The general quantum interference principle and the
duality computer.} Commun. Theor. Phys. {\bf 45}, 825-844 (2006) .


\bibitem{r2} Wang, W Y, Shang, B, Wang, C and Long, G L, {Prime factorization
in the duality computer.} Commun. Theor. Phys. {\bf 47}, 471-473
(2007).

\bibitem{r3} Gudder, S, {Duality quantum computers.}
Quantum Information Processing, \textbf{6} (1): 49-54, (2007).

\bibitem{r4}  Long, G L, {Mathematical theory of duality computer in the density matrix
formalism.} Quantum Information Processing, \textbf{6}(1): 49-54,
(2007)

\bibitem{r5} Long, G L and Liu, Y,
Search an unsorted database with quantum mechanics.
 Front. Comput.
Sci. China, \textbf{1} (3): 247-271 (2007)


\bibitem{r7} Brassard, G, Hoyer, P, Mosca, M and Tapp, A ,
Quantum amplitude amplification and esti-mation. AMS Contemporary
Mathematics Series Vol.\textbf{305}, eds. S. J. Lomonaco and H. E.
Brandt, AMS (Providence), p.53, (2002)

\bibitem{r8} Hoyer, P,  Arbitrary phases in quantum amplitude
amplification.  Phys. Rev. A \textbf{62}, 052304 (2001).

\bibitem{r9} Long, G L, {Grover algorithm with zero failure rate,} Phys. Rev.
A64, 022307 (2001).

\bibitem{r10} Grover, L K, Fixed-point quantum search, Phys. Rev. Lett.
\textbf{95},
 150501 (2005).

\bibitem{r11} Shor, P W, Scheme for reducing decoherence in quantum computer
memory.
 Phys. Rev. A. \textbf{52}, R2493 (1995).


\end{thebibliography}
\end{document}